\documentstyle[12pt]{article}

\def\pd{\partial}
\def\a{\alpha}

\def\dl{\delta}
\def\s{\sigma}

\def\lam{\lambda}
\def\Lam{\Lambda}
\def\bg{{\bar g}}
\def\hg{{\hat g}}

\def\bnabla{{\bar \nabla}}
\def\hnabla{{\hat \nabla}}
\def\bR{{\bar R}}

\def\bBox{\stackrel{-}{\Box}}
\def\hBox{{\hat \Box}} 

\def\gm{\gamma}

\def\sq{\sqrt}
\def\e{\hbox{\large \it e}}
\def\half{\frac{1}{2}}
\def\fr{\frac}
\def\pp{\prime}

\def\bb{\begin{equation}}
\def\ee{\end{equation}}
\def\bba{\begin{eqnarray}}
\def\eea{\end{eqnarray}}

\begin{document}

\begin{titlepage}

\begin{tabbing}
   qqqqqqqqqqqqqqqqqqqqqqqqqqqqqqqqqqqqqqqqqqqqqq 
   \= qqqqqqqqqqqqq  \kill 
         \>  {\sc KEK-TH-687}    \\
         \>  {\sc May, 2000} 
\end{tabbing}
\vspace{1.5cm}

\begin{center}
{\Large {\bf On The BRST Formulation 
of Diffeomorphism Invariant \\ 4D Quantum Gravity}}
\end{center}

\vspace{1.5cm}

\centering{\sc Ken-ji Hamada\footnote{E-mail address : 
hamada@post.kek.jp} }

\vspace{1cm}

\begin{center}
{\it Institute of Particle and Nuclear Studies, \break 
High Energy Accelerator Research Organization (KEK),} \\ 
{\it Tsukuba, Ibaraki 305-0801, Japan}
\end{center} 

\vspace{1.5cm}

\begin{abstract} 
In this note we give some remarks on the BRST formulation of 
a renormalizable and diffemorphism invariant 4D quantum gravity 
recently proposed by the author, which satisfies the integrability 
condition by Riegard, Fradkin and Tseytlin at the 2-loop level. 
Diffeomorphism invariance requires an addition of the Wess-Zumino 
action, from which the Weyl action can be induced by expanding around 
a vacuum expectation value of the conformal mode. 
This fact suggests the theory has in itself a mechanism to remove 
extra negative-metric states dynamically. 


\end{abstract}
\end{titlepage}  
 
 Diffeomorphism invariance requires that 4D quantum gravity 
becomes 4th order derivative theory for gravity sector. 
We recently showed~\cite{h1,hs} that 4th order  actions, 
including the Wess-Zumino (WZ) action~\cite{r,amm},  
are uniquely determined by diffeomorphism invariance. 
Then, the theory also becomes renormalizable~\cite{h1}. 
Especially, our model satisfies the 
integrability condition on the WZ action discussed 
by Riegard, Fradkin and Tseytlin~\cite{r}, 
which is generalized by the author  
to the form that can be applied to higher loops.  
A problem in 4th-order theories is that there are extra 
negative-metric states. 
Thus, the unitarity becomes obscure~\cite{t}.    
In this paper we shall see that there is a posibility that 
diffeomorphism invariance also ensures the unitarity. 

  Here, we briefly explain how to realize diffeomorphism invariance.    
The details of the argument were discussed in our previous 
papers~\cite{h1,hs}.  
Perturbation theory is defined by replacing the invariant measure 
with the measure defined on the background-metric. 
As a lesson from 2D quantum gravity~\cite{kpz}--\cite{kkn}, 
in order to preserve background-metric independence, 
or diffeomorphism invariance, we must add an action, $S$, 
which satisfies the WZ condition~\cite{wz}, as 
\bb
     Z= \int \frac{[d\phi]_{\hg} [\e^{-h}d \e^h]_{\hg} 
                    [df]_{\bg} }{\hbox{vol(diff.)}} 
                 \exp \bigl[-S(\phi,\bg) - I(f,g) \bigr] ~,
                    \label{zz}
\ee
where $f$ is a matter field and $I$ is an invariant action. 
The metric is now decomposed as $g_{\mu\nu}=\e^{2\phi}\bg_{\mu\nu}$ 
and $\bg_{\mu\nu}=(\hg \e^h)_{\mu\nu}$, 
where $tr(h)=0$~\cite{kkn}.   
The measures of the metric fields are defined on the background-metric 
by the norms:
\bba
    && <d \phi, d \phi>_{\hg} 
        = \int d^4 x \sq{\hg} (d \phi)^2 ~,
                    \\
    && <d h, d h>_{\hg} 
        = \int d^4 x \sq{\hg} ~tr (\e^{-h} d \e^h)^2 ~. 
\eea

  The general coordinate transformation, 
$\dl g_{\mu\nu} =g_{\mu\lam}\nabla_{\nu}\xi^{\lam} 
+g_{\nu\lam}\nabla_{\mu}\xi^{\lam}$,  
is expressed in 4 dimensions as  
\bba
      && \dl \phi = \frac{1}{4} \hnabla_{\lam}\xi^{\lam} +  
                     \xi^{\lam}\pd_{\lam}\phi ~, 
                      \nonumber   \\ 
      && \dl \bg_{\mu\nu} = \bg_{\mu\lam}\bnabla_{\nu}\xi^{\lam} 
                 +\bg_{\nu\lam}\bnabla_{\mu}\xi^{\lam}  
                 -\frac{1}{2}\bg_{\mu\nu}\hnabla_{\lam}\xi^{\lam} ~,    
                      \label{gct}
\eea  
where $\bnabla_{\lam} \xi^{\lam}= \hnabla_{\lam} \xi^{\lam}$ is used.
Under a general coordinate transformation, $\dl I=0$, 
but the WZ action is not invariant. $\dl S$ is proportional to 
the form of {\it conformal anomaly}~\cite{duff}.   
Diffeomorphism invariance is realized such that  
$\dl S$ cancels anomalous contributions, $U$, which originates from 
the fact that the measures defined above are no longer invariant 
under the transformation, as 
\bb
     \dl Z = - < \dl S + U>=0 ~. 
\ee  
More rigorously, consider the regularized 1PI effective action, 
$\Gamma_{\rm eff}$, of the combined theory, ${\cal I}=S+I$, 
and require $\dl \Gamma_{\rm eff}=0$, which determines 
$S$ uniquely.

    In 2 dimensions we can take the conformal gauge $h^{\mu}_{~\nu}=0$, 
and hence 2D quantum gravity coupled to conformal matter can be 
described as a free conformal field theory~\cite{dk}.  
Of course, in 4 dimensions, the combined theory ${\cal I}=S+I$ can not 
be described as a free theory.
We must take into account interactions between 
the conformal mode and the traceless mode in the WZ action 
as well as self-interactions of the traceless mode, 
which are ruled by the background-metric independence for  
the traceless mode~\cite{hs,h1}.  
Thus, we must generalize the idea of 2D quantum gravity 
based on conformal field theories to one based 
on diffeomorphism invariance.  
The original idea on this matter is given in 
a study of 2D quantum dilaton 
gravity~\cite{h3,ht},\footnote{ 
There is an analogy between the dilaton field $\varphi$ defined 
in~\cite{h3,ht} and the traceless mode in our 4D model.  
Unfortunately, this model is unrenormalizable in the perturbation 
of non-minimal coupling because $\varphi$ is a dimensionless 
scalar in 2 dimensions so that there are many diffeomorphism 
invariant counterterms like  
$\varphi^n \pd^{\mu}\varphi\pd_{\mu}\varphi$.
On the other hand the dynamics of 
the traceless mode is ruled by the background-metric independence 
for the traceless mode, itself, so that the model has almost no 
ambiguity.
} 
and then developed to 4D quantum gravity~\cite{hs}.   

  The conformal-mode dynamics of the WZ action in 4 dimensions has been  
discussed in refs.~\cite{amm} in analogy to 2D quantum gravity. 
But, in their model there is no interactions for the traceless 
mode,  because they considered the WZ action as a full effective 
action given after the traceless mode as well as matter fields 
are integrated out. From the viewpoint of full diffeomorphism 
invariance, it is inaccurate in 4 dimensions. 
 
 In this note we give the BRST formulation~\cite{brs,kugo} 
of diffeomorphism invariant quantum gravity. 
We first review the BRST formulation of 
2D quantum gravity~\cite{kato,itoh}.  
We here emphasize that the nilpotence of 
the BRST transformation, which is equivalent to 
diffeomorphism invariance, is realized dynamically.  
We then show that similar considerations can apply to diffeomorphism 
invariant 4D quantum gravity.  
At the end we discuss the possibility of how to remove 
the negative-metric states in the 4D model  
from the viewpoint of diffeomorphism invariance.  
Preliminary discussions on this matter have already 
given in~\cite{h1}. 

Our curvature conventions are $R_{\mu\nu}= R^{\lam}_{~\mu\lam\nu}$
and $R^{\lam}_{~\mu\s\nu}= \pd_{\s} \Gamma^{\lam}_{~\mu\nu} - \cdots$. 

\begin{flushleft}
{\bf \underline{2D quantum gravity}}
\end{flushleft}  
  
  Firstly, we briefly review the BRST formulation of 
2D quantum gravity. The WZ action in two dimensions, 
what is called the Liouville action, is given by integrating 
the 2D conformal anomaly as~\cite{p}
\bb
    S(\phi,\bg) = \fr{a}{4\pi} \int d^2 x \sq{\hg} 
    \bigl( \bg^{\mu\nu}\pd_{\mu} \phi \pd_{\nu} \phi 
     + \bR \phi \bigr) ~. 
\ee
In 2 dimensions we can take the gauge 
condition $h^{\mu}_{~\nu}=0$ up to the zero mode. 
The gauge-fixed combined action then becomes 
\bb
     {\cal I} = \frac{1}{4\pi}\int d^2 x \sq{\hg} \biggl[ 
             a  \Bigl( \bg^{\mu\nu} \pd_{\mu}\phi \pd_{\nu}\phi 
             + {\bar R}\phi \Bigr)   
             + {\cal L}_{GF+FP} + \Lam \e^{\a\phi} \biggr] 
                  + I_M(f,g) ~,
               \label{gfa}
\ee  
where $I_M$ is an invariant matter action. The gauge-fixing term 
and the Faddeev-Popov (FP) ghost action are given by
\bb
      {\cal L}_{GF+FP} 
       = -i B_{\mu\nu}(\bg^{\mu\nu} -\hg^{\mu\nu}) 
         +2 \bg^{\mu\nu} b_{\mu\lam} \bnabla_{\nu} c^{\lam} ~,  
\ee
where the reparametrization ghost $c^{\mu}$ is a contravariant vector.  
$B_{\mu\nu}$ and the anti-ghost $b_{\mu\nu}$ are covariant symmetric 
traceless tensors. 
In the following, ${\cal I}$ is considered as a "classical" action.

  Consider $N$ massless scalars as matter fields.  
Then, diffeomorphism invariance requires that the coefficient, 
$a$, must be~\cite{dk}  
\bb
    a = \fr{1}{6}(25-N) ~. \label{a}
\ee 
 
  The BRST transformation is given by 
\bba
    && {\bf \dl_B} \bg_{\mu\nu} 
       = i ( \bg_{\mu\lam}\bnabla_{\nu} c^{\lam} 
              +\bg_{\nu\lam}\bnabla_{\mu} c^{\lam}  
             -\bg_{\mu\nu}\hnabla_{\lam} c^{\lam}) ~,  
                 \nonumber       \\ 
    && {\bf \dl_B} \phi 
       = i c^{\lam} \pd_{\lam} \phi  
         + i\frac{1}{2} \hnabla_{\lam} c^{\lam} ~, 
                 \nonumber       \\ 
    && {\bf \dl_B} b_{\mu\nu} = B_{\mu\nu} ~, \qquad 
       {\bf \dl_B} B_{\mu\nu} = 0 ~, 
               \label{brst2a}         \\ 
    && {\bf \dl_B} c^{\mu} 
        = i c^{\lam}\pd_{\lam} c^{\mu} ~,  
               \nonumber
\eea 
where $c^{\lam}\bnabla_{\lam} c^{\mu}=c^{\lam}\pd_{\lam} c^{\mu}$ 
due to the anti-commutativity of $c^{\mu}$. 
Note that the $h$-dependence appears only in $\dl_B \bg_{\mu\nu}$. 
These transformations are nilpotent; 
$\dl_B^2 (\bg_{\mu\nu},~\phi,~b_{\mu\nu},~c^{\mu})=0$.  
Then, the gauge-fixing term and the FP ghost action can be 
written as ${\cal L}_{GF+FP}= -i{\bf \dl_B} 
\{ b_{\mu\nu}(\bg^{\mu\nu} -\hg^{\mu\nu}) \}$~\cite{ku}, 
where $\dl_B \bg^{\mu\nu}=-\bg^{\mu\lam}\bg^{\nu\s}\dl_B \bg_{\lam\s}$.  
    
 The well-known form of the BRST transformation in 2D quantum gravity 
is given by integrating out over the $B_{\mu\nu}$ field. 
Hence, we obtain the following one:
\bba
    && {\bf \dl_B} \phi 
       = i c^{\lam} \pd_{\lam} \phi  
         + i\frac{1}{2} \hnabla_{\lam} c^{\lam} ~, 
                \nonumber        \\ 
    &&  {\bf \dl_B} b_{\mu\nu} = 2 i{\cal T}_{\mu\nu} ~,
                 \label{brst2b}       \\ 
    && {\bf \dl_B} c^{\mu} 
        = i c^{\lam}\pd_{\lam} c^{\mu} ~.  
                 \nonumber
\eea 
Now, $h^{\mu}_{~\nu}$ becomes non-dynamical and we can set 
$\bg_{\mu\nu}=\hg_{\mu\nu}$ in the expressions. 
To guarantee $\dl_B h^{\mu}_{~\nu}=0$,  
$c^{\mu}$ should satisfy the conformal Killing equation, 
$\hnabla^{\mu} c^{\nu} + \hnabla^{\nu} c^{\mu}  
-\hg^{\mu\nu}\hnabla_{\lam} c^{\lam} = 0$.    
${\cal T}_{\mu\nu}$ is the modified energy-momentum tensor, 
which is determined by using the equation of motion for the traceless 
mode and tracelessness of $b_{\mu\nu}$, 
as~\footnote{ 
Because of the tracelessness of $b_{\mu\nu}$, 
there is an ambiguity $\gm$, which is    
that one can add  $\gm b_{\mu\nu}\hnabla_{\lam}c^{\lam}$ 
to the energy-momentum tensor for the $bc$-system.  
It is now fixed by the condition $\dl_B {\cal I}_{bc}=0$.   
} 
\bba
   {\cal T}_{\mu\nu} 
      &=& T_{\mu\nu} - \half \hg_{\mu\nu}T^{\lam}_{~\lam}  
             \nonumber    \\ 
      &=& -\frac{a}{2} \biggl\{ 
            \pd_{\mu}\phi \pd_{\nu}\phi 
            -\half \hg_{\mu\nu} \pd^{\lam}\phi \pd_{\lam}\phi 
            -\Bigl( \hnabla_{\mu}\hnabla_{\nu} 
                -\half \hg_{\mu\nu} \hBox \Bigr) \phi \biggr\} 
             \nonumber  \\ 
      && + \hnabla_{(\mu} c^{\lam} b_{\nu) \lam} 
         + \half c^{\lam} \hnabla_{\lam} b_{\mu\nu} 
            -\half \hg_{\mu\nu} \hnabla^{\lam}c^{\s} b_{\lam\s} 
         + T^M_{\mu\nu} ~. 
              \label{emt}
\eea
Here, $T_{\mu\nu} = -\frac{2\pi}{\sq{\hg}}
\frac{\dl{\cal I}}{\dl \hg^{\mu\nu}}$ 
is the energy-momentum tensor of the gauge-fixed combined theory 
(\ref{gfa}) with $h^{\mu}_{~\nu}=0$. 
$T^M_{\mu\nu}$ is the energy-momentum tensor of $N$ massless scalars.  
Since, in two dimensions, ${\cal T}^{\lam}_{~\lam}=0$, 
2D quantum gravity can be expressed by conformal field theory.
 
  The transformation (\ref{brst2b}) is now no longer 
nilpotent "classically".  Using expression (\ref{emt}) and 
the conformal Killing equation of $c^{\mu}$,  
we can show that $\dl_B^2 b_{\mu\nu}$ produces a non-vanishing 
quantity,  
\bb
    \dl_B {\cal T}_{\mu\nu}
    =i\frac{a}{4} \biggl( \hnabla_{\mu}\hnabla_{\nu} 
      -\half \hg_{\mu\nu} \hBox \biggr) \hnabla_{\lam}c^{\lam} ~,   
\ee 
which implies that the energy-momentum tensor, 
${\cal T}_{\mu\nu}$, forms Virasoro algebra with central 
charges $6a$ classically~\cite{ht}.  
It reflects that the action, ${\cal I}$, is not 
BRST-invariant:~\footnote{ 
Eq. (\ref{var2d}) can be derived either by applying (\ref{brst2a}) 
to the action (\ref{gfa}) 
or by applying (\ref{brst2b}) to the action obtained by integrating the 
$B_{\mu\nu}$ field out, where the conformal Killing equation of 
$c^{\mu}$ is not necessary even in the later case.
} 
\bb
    \dl_B {\cal I}= \frac{ia}{8\pi} 
    \int d^2 x \sq{\hg} \bR \hnabla_{\lam} c^{\lam} ~. 
            \label{var2d}
\ee       

  The BRST invariance of 2D quantum gravity is realized dynamically 
as follows. 
Quantum effects, namely, conformal anomalies give contributions 
to the central charge, $N-25$,    
so that the total of the central charge becomes $c_{tot}=6a+N-25$.     
Thus, the nilpotence of the BRST transformation 
is realized at the quantum level when $a$ is given by 
equation (\ref{a}).    

  Finally, we give  brief comments on physical states 
in 2D quantum gravity. 
The presence of the $\half \hnabla_{\lam} c^{\lam}$ term 
in the BRST transformation of $\phi$  
implies that the asymptotic state of the conformal mode, $\phi|0>$, 
is not physical~\cite{kugo}. On the other hand, 
the BRST invariant state, naively defined by $ \e^{\a \phi}|0>$, 
where $\a$ is a real value determined by the BRST invariance, is not 
normalizable~\cite{s}. Furthermore, in 2 dimensions, the normalizable 
Hamiltonian eigenstates are not BRST invariant because they clearly do 
not satisfy the Hamiltonian constraint, $H={\cal T}^0_{~0} =0$. 
Thus, there is no gravitational degrees of freedom in 2D quantum gravity.  

\begin{flushleft}
{\bf \underline{4D quantum gravity}}
\end{flushleft} 
 
  In 4 dimensions the WZ action becomes 4th order, 
parametrized by three constants $a$, $b$ and $c$ in the form
\bba
    S(\phi,\bg) 
        &=& \frac{1}{(4\pi)^2} \int d^4x \biggl[ \sq{\hg} 
            \biggl\{ a {\bar F} \phi 
            +2b \phi {\bar \Delta}_4 \phi 
            +b \Bigl( {\bar G}-\fr{2}{3} 
            \bBox {\bar R} \Bigr) \phi \biggr\}            
                  \nonumber \\ 
        && \qquad\qquad\qquad
          - \frac{1}{36}(2a+2b+3c) 
          \Bigl( \sq{g} R^2 
          -\sq{\hg}{\bar R}^2 \Bigr) \biggr] ~,   
\eea
where $F$ is the square of the Weyl tensor and $G$ is the Euler 
density defined by 
\bba
    F &=& R_{\mu\nu\lam\s}R^{\mu\nu\lam\s}-2R_{\mu\nu}R^{\mu\nu} 
           +\fr{1}{3}R^2 ~,
               \\ 
    G &=& R_{\mu\nu\lam\s}R^{\mu\nu\lam\s}-4R_{\mu\nu}R^{\mu\nu} 
           +R^2 ~. 
\eea   
$\Delta_4$ is the conformally covariant 4th 
order operator~\cite{r}, 
\bb
      \Delta_4 = \Box^2 
               + 2 R^{\mu\nu}\nabla_{\mu}\nabla_{\nu} 
                -\fr{2}{3}R \Box 
                + \fr{1}{3}(\nabla^{\mu}R)\nabla_{\mu} ~. 
\ee
Why the number of the independent parameters is three 
is due to the fact that $R^2$ is not integrable w.r.t. 
conformal mode~\cite{r}. 
 
   We consider the following invariant action including 
4th order terms:
\bb
     I(f,g)=I_4 +I_{LE} ~,  \label{cla}
\ee
where  
\bba
     && I_4 = \fr{d}{(4\pi)^2} \int d^4 x \sq{g} R^2  ~, 
                      \\ 
     &&  I_{LE} = \fr{1}{(4\pi)^2} \int d^4 x \sq{g} 
                   (- m^2 R + \Lambda ) + I_M (f,g) ~.    
\eea
$I_4$ is the 4th order action with $d>0$ which as well as 
the WZ action could be regarded 
as being a part of the measure. $I_{LE}$ is the usual 2nd order 
action which describes low-energy physics. 
$m^2$ is the inverse of the gravitational constant 
and $\Lambda$ is the cosmological constant. 
$I_M$ is a matter action. 
Here, note that the presence of the $\phi{\bar F}$ term 
in the WZ action implies that  
the Weyl term, $\sq{g}F ~(=\sq{\hg}{\bar F})$, can be produced  
by expanding around a vacuum expectation value (VEV) of $\phi$. 

  Let us define the perturbation around VEV of $\phi$ such that 
the Weyl term, $\frac{1}{t^2}{\bar F}$, is produced, where we 
introduce the dimensionless coupling constant, $t$, for the 
traceless mode and the $h^{\mu}_{~\nu}$ field is replaced with 
$t h^{\mu}_{~\nu}$ in the combined action~\cite{hs,h1}. 
Then, the integral region of $\phi$ is effectively restricted within the 
region $-\frac{1}{t} < \phi <\frac{1}{t}$. This perturbation theory 
seems to be well-defined; namely, it is expected that $t$ is small 
enough~\cite{h1} in comparison with recent numerical 
experiments~\cite{bbkptt}.

  Let us consider the regularized 1PI effective action, 
$\Gamma_{\rm eff}$. As discussed in~\cite{h1}, diffeomorphism 
invariance, namely, $\dl \Gamma_{\rm eff}=0$ gives constraints on 
actions of gravitational fields as well as matter fields,    
which requires that 
4D model must satisfy the following conditions:    
\begin{itemize}
\item Matter fields must {\it conformally} couple to gravity. 
\item The coefficient, $d$, in 4th order action must be  
\bb 
     d=\frac{1}{36}(2a+2b+3c) ~. \label{d}
\ee   
\end{itemize}   
The second condition means that self-interactions of the 
conformal mode, namely the $R^2$ terms, cancel out in the 
combined action. Then, the three coefficients $a$, $b$ and $c$, 
can be determined uniquely in the perturbation of the coupling, $t$, 
by requiring diffeomorphism 
invariance.\footnote{
The coefficients depend on matter contents, but the sign of $a$ 
($b$) is negative-definite (positive-definite) at the zero-th order 
of $t$. $c$ vanishes at this order. And also the sum of them, $d$, 
becomes positive. 
} 

  These conditions are more precisely represented as that,  
in the regularized 1PI effective action, 
there is no non-local correction to the WZ action 
like $\phi \bBox^2 \log (-\bBox) \phi$  
and no non-local term, $\bR \log (-\bBox) \bR$, which is associated to 
non-conformally invariant counterterm proportional 
to $\bR^2$.   
This is the generalized form of the integrability condition 
discussed in~\cite{r}.
Here, there are two remarks. The first is that the two types of 
non-local corrections considered here 
are related to each other by the background-metric 
independence for the conformal mode. 
The second is that vanishing of the $\bR \log (-\bBox) \bR$ term 
does not always imply vanishing of the $\bR^2$ divergences 
at higher loops. 

 The combined action, ${\cal I} =S+I$, including the gauge-fixing 
term and the FP ghost action, now becomes
\bba
   &&{\cal I}= \frac{1}{(4\pi)^2} \int d^4 x 
        \biggl[ ~\frac{1}{t^2}{\bar F} + a {\bar F} \phi 
          + 2 b \phi  {\bar \Delta}_4 \phi 
          + b \Bigl( {\bar G}-\fr{2}{3} \bBox {\bar R} \Bigr) \phi 
              \nonumber   \\
   && \qquad\qquad\quad
        + \frac{1}{36}(2a+2b+3c){\bar R}^2 
        + {\cal L}_{GF+FP}  \biggr] + I_{LE}(f,g) ~.
                \label{act}
\eea 
Here and below, we take the flat background $\hg_{\mu\nu}=\dl_{\mu\nu}$. 
The gauge-fixing term is given by~\cite{bjs}.  
\bb
      {\cal L}_{GF}  = 2i B^{\mu} N_{\mu\nu} \chi^{\nu}   
            - \zeta B^{\mu} N_{\mu\nu} B^{\nu}  ~,  
                         \label{gfix}
\ee
where $\chi^{\nu}=\pd^{\lam}h^{\nu}_{~\lam}$ and $N_{\mu\nu}$ is a 
symmetric 2nd order operator. 
The corresponding ghost action becomes 4th order: 
\bb
     {\cal L}_{FP} = -2i {\bar c}^{\mu} N_{\mu\nu} \pd^{\lam} 
              {\bf \dl_B} h^{\nu}_{~\lam} ~,
\ee
where the BRST transformation of the traceless mode is given by 
\bba
      {\bf \dl_B} h^{\mu}_{~\nu}  
      &=& i \biggl[ \pd^{\mu} c_{\nu} 
                       +\pd_{\nu} c^{\mu}
           - \half \dl^{\mu}_{~\nu}  
                    \pd_{\lam} c^{\lam}   
           + t c^{\lam} \pd_{\lam} h^{\mu}_{~\nu}   
                \nonumber    \\ 
       &&
           + \frac{t}{2} h^{\mu}_{~\lam} 
                \Bigl( \pd_{\nu} c^{\lam} 
                  - \pd^{\lam} c_{\nu} \Bigr) 
           + \frac{t}{2} h^{\lam}_{~\nu} 
                \Bigl( \pd^{\mu} c_{\lam} 
                  - \pd_{\lam} c^{\mu} \Bigr) 
           + \cdots \biggr]  ~.
                   \label{brst4a} 
\eea
This is given by replacing $\xi^{\mu}/t$ in the transformation with 
the contravariant vector ghost field, $ic^{\mu}$.  
The kinetic term of the ghost  
action then becomes $t$-independent. 
The BRST transformations for other fields are given by
\bba
    && {\bf \dl_B} \phi 
       = i t c^{\lam} \pd_{\lam} \phi 
         + i\frac{t}{4} \pd_{\lam} c^{\lam} ~, 
              \nonumber          \\ 
    && {\bf \dl_B} {\bar c}^{\mu} =  B^{\mu} ~,  \qquad
       {\bf \dl_B} B^{\mu} = 0 ~, 
             \label{brst4b}           \\ 
    && {\bf \dl_B} c^{\mu} 
       = i t c^{\lam}\pd_{\lam} c^{\mu}  ~.   
            \nonumber
\eea
The transformations, (\ref{brst4a}) and (\ref{brst4b}), are nilpotent. 
Using the BRST transformation, the gauge-fixing term and the FP ghost 
action can be written as ${\cal L}_{GF+FP}= 2i{\bf \dl_B} 
\{ {\bar c}^{\mu}N_{\mu\nu} (\chi^{\nu} 
+ \frac{i}{2}\zeta B^{\nu}) \}$~\cite{ku}.  

  As in 2D quantum gravity, the BRST transformation of the "classical" 
action, ${\cal I}$, is not BRST-invariant:  
\bb
     \dl_B {\cal I} = \fr{it}{4(4\pi)^2} \int d^4 x
                      \pd_{\lam} c^{\lam} 
                \biggl[ a \biggl( {\bar F} + \fr{2}{3}\bBox \bR 
                           \biggr) +b {\bar G} 
                           + c \bBox \bR \biggr] ~. 
\ee
The BRST invariance is equivalent to diffeomorphism invariance,  
which is realized dynamically as mentioned before. 
  
  If the $B^{\mu}$ field is integrated out, 
this field is related to the energy-momentum tensor through the field 
equation for the traceless mode.  
As in 2D quantum gravity, it means that the nilpotence of 
the BRST transformation requires that the BRST transformation of the 
energy-momentum tensor vanishes at the quantum level.
  
  Let us consider the long-distance physics of this model. 
The theory is asymptotically free, namely, $a < 0$  
for the coupling of the traceless mode, $t$.  
Thus, one can drop the Weyl term at the long distance. 
On the other hand, we leave the other three kinetic terms 
of gravitational fields: 
$\phi {\bar \Delta}_4 \phi$, $\bR^2$ and $m^2 R$.
Let us compute the degrees of freedom in this case.  
Since $\bR^2= -\chi^{\mu}\pd_{\mu}\pd_{\nu}\chi^{\nu}+o(h^3)$, 
the 2nd-order operator, $N_{\mu\nu}$, is proportional to 
$-\pd_{\mu}\pd_{\nu} + m^{\pp 2} 
\dl_{\mu\nu}$,\footnote{ 
Although there are the off-diagonal $\phi-h $ terms 
in the $\phi \bBox\bR$ term and the Einstein-Hilbert term,  
they can be diagonalized into a 4th-order kinetic term of $\phi$ 
and this form for $h$, provided that one set $t=1$.     
} 
where $m^{\pp 2}$ comes from the Einstein-Hilbert term, which is 
proportional to $m^2$. The ghost determinant is now given by 
$\det^{1/2} (-\pd_{\mu}\pd_{\nu}+m^{\pp 2}\dl_{\mu\nu}) 
\det M^{GH}_{\mu\nu}$, 
where $M^{GH}_{\mu\nu}$ is the usual 2nd order ghost operator 
of diffeomorphism invariance given by applying 
the BRST transformation to $\chi^{\mu}$ 
such that $\dl_B \chi_{\mu} |_{h=0}=M^{GH}_{\mu\lam}c^{\lam}$.  
Note that 
$\det (-\pd_{\mu}\pd_{\nu}+m^{\pp 2}\dl_{\mu\nu}) 
= \det (-\Box +m^{\pp 2}) \vert_{\rm a ~scalar}$, 
so that the number of ghost degrees of freedom is $4+4+1=9$. 
Hence, the total degrees of freedom becomes  $2 \times 1+9-9=2$.  
Thus, the negative-metric state related to the conformal mode is   
removed by ghosts. Really, due to the form of the BRST transformation, 
the conformal mode can not be a BRST invariant 
asymptotic state~\cite{kugo}. 
  
   This result is quite suggestive, because, if there is a mechanism 
to remove the Weyl term, it seems that the theory becomes unitary.   
Recall that an exact diffeomorphism invariance implies that the 
integral region of $\phi$ is unrestricted above and below, so that 
the Weyl term can be absorbed into the $\phi {\bar F}$ term in 
the WZ action, which is just the original theory (\ref{cla}) 
with (\ref{d}) adding the WZ action. 
Thus, we expect that diffeomorphism invariance 
ensures the unitarity non-perturbatively. 

\vspace{5mm}

\begin{flushleft}
{\bf Acknowlegments}
\end{flushleft}

  This work is supported in part by the Grant-in-Aid for 
Scientific Research from the Ministry of 
Education, Science and Culture of Japan.

\end{document}